\documentclass{llncs}
\usepackage{llncsdoc}
\usepackage{epsf}
\usepackage{amsfonts}
\usepackage{tabularx}
\usepackage{mathrsfs}
\usepackage{amsmath}
\usepackage{multirow}
\usepackage{graphicx}
\usepackage{mathtools}

\newcommand\nc{\newcommand}

\def\qed{\rule{1.5mm}{3mm}}

\nc{\crl}[2]{\begin{corollary}\label{crl:#1} #2 \end{corollary}}
\nc{\lem}[2]{\begin{lemma}\label{lem:#1} #2 \end{lemma}}
\nc{\prp}[2]{\begin{proposition}\label{prp:#1} #2 \end{proposition}}
\nc{\thm}[2]{\begin{theorem}\label{thm:#1} #2 \end{theorem}}
\nc{\cnj}[2]{\begin{conjecture}\label{cnj:#1} #2 \end{conjecture}}
\nc{\que}[2]{\begin{question}\label{que:#1} #2 \end{question}}
\nc{\pro}[2]{\begin{problem}\label{pro:#1} #2 \end{problem}}
\nc{\dfn}[2]{\begin{definition}\label{def:#1} #2 \end{definition}}

\nc{\llem}[3]{\begin{lemma}[#2]\label{lem:#1} #3 \end{lemma}}
\nc{\tthm}[3]{\begin{theorem}[#2]\label{thm:#1} #3 \end{theorem}}
\nc{\ddfn}[3]{\begin{definition}[#2]\label{def:#1} #3 \end{definition}}

\nc{\fig}[4]{\begin{figure}[hbt]
\centerline{
\epsfysize=#2in
\epsffile{#4}
}
\caption{#3}
\label{fig:#1}
\end{figure}}

\nc{\tbl}[3]{\begin{table}[hbt]
#3
\caption{#2}
\label{tab:#1}
\end{table}}

\nc{\refc}[1]{Corollary~\ref{crl:#1}}
\nc{\reff}[1]{Figure~\ref{fig:#1}}
\nc{\refj}[1]{Conjecture~\ref{cnj:#1}}
\nc{\refl}[1]{Lemma~\ref{lem:#1}}
\nc{\refp}[1]{Proposition~\ref{prp:#1}}
\nc{\reft}[1]{Theorem~\ref{thm:#1}}
\nc{\refq}[1]{Question~\ref{que:#1}}
\nc{\refb}[1]{Problem~\ref{pro:#1}}
\nc{\reffct}[1]{Fact~\ref{fct:#1}}
\nc{\reftb}[1]{Table~\ref{tab:#1}}

\begin{document}

\title{ 
\Large \bf  Finding Two Edge-Disjoint Paths with Length Constraints
}
\author{
	Leizhen CAI\thanks{Partially supported by GRF grant CUHK410212 of the Research Grants Council of Hong Kong.} and Junjie YE}
\institute{Department of Computer Science and Engineering, The Chinese University of Hong Kong, Shatin, Hong Kong SAR, China \\
	\email{\{lcai,jjye\}@cse.cuhk.edu.hk}
}
\maketitle

\begin{abstract}
	We consider the problem of finding, for two pairs $(s_1,t_1)$
	and $(s_2,t_2)$ of vertices in an undirected graphs, an $(s_1,t_1)$-path $P_1$
	and an $(s_2,t_2)$-path $P_2$ such that $P_1$ and $P_2$ share no edges and
	the length of each $P_i$ satisfies $L_i$, where
	$L_i \in \{ \le k_i, \; = k_i, \; \ge k_i, \; \le \infty\}$.
	
	We regard $k_1$ and $k_2$ as parameters and investigate the parameterized complexity
	of the above problem when at least one of $P_1$ and $P_2$ has a length constraint
	(note that $L_i = ``\le \infty"$ indicates that $P_i$ has no length constraint).
	For the nine different cases of $(L_1, L_2)$, we obtain FPT algorithms for seven of them.
	Our algorithms uses random partition backed by some structural results.
	On the other hand,  we prove that the problem admits no polynomial kernel for
	all nine cases unless $NP \subseteq coNP/poly$. \\

	\noindent {\bf Keywords:} Edge-disjoint paths, random partition, parameterized complexity,
	kernelization.
\end{abstract}

\section{Introduction}

Disjoint paths in graphs are fundamental and have been studied extensively in the literature. 
Given $k$ pairs of {\em terminal vertices} $(s_i, t_i)$ for $1 \le i \le k$ in an undirected 
graph $G$, the classical {\sc Edge-Disjoint Paths} problem asks whether $G$ contains $k$ 
pairwise edge-disjoint paths $P_i$ between $s_i$ and $t_i$ for all $1 \le i \le k$. 
The problem is NP-complete as shown by Itai et al.~\cite{itai1982complexity},
but is solvable in time $O(mn)$ by network flow~\cite{orlin2013max} 
if all vertices $s_i$ (resp.,  $t_i$) are the same vertex $s$ (resp.,  $t$). 
When we regard $k$ as a parameter, a celebrated result of Robertson and 
Seymour~\cite{robertson1995graph} on vertex-disjoint paths can be used to
obtain an FPT algorithm for {\sc Edge-Disjoint Paths}.
On the other hand, Bodlaender et al.~\cite{bodlaender2011kernel} have shown that 
{\sc Edge-Disjoint Paths} admits no polynomial kernel unless $NP \subseteq coNP/poly$.  

In this paper, we study {\sc Edge-Disjoint Paths} with length constraints $L_i$ on 
$(s_i,t_i)$-paths $P_i$ and focus on the problem for two pairs of terminal vertices.
The length constraints $L_i \in \{ \le k_i, \; = k_i, \; \ge k_i, \; \le \infty\}$ indicate
that the length of $P_i$ need to satisfy $L_i$.
We regard $k_1$ and $k_2$ as parameters, and study the parameterized complexity 
of the following problem.

\begin{quote}
	{\sc Edge-Disjoint $(L_1,L_2)$-Paths}
	
	{\sc Instance}: Graph $G = (V,E)$, two pairs $(s_1,t_1)$ and $(s_2,t_2)$ of vertices.
	
	{\sc Question}: Does $G$ contain $(s_i,t_i)$-paths $P_i$ for $i = 1,2$ 
	such that $P_1$ and $P_2$ share no edge and the length of $P_i$ satisfies $L_i$?
\end{quote}

There are nine different length constraints on two paths 
(note that {\sc Edge-Disjoint $(\le \infty, \le \infty)$-Paths} puts no length constraint on two paths). 
For instance, {\sc Edge-Disjoint $(= k_1, \le \infty)$-Paths}
requires that $|P_1| = k_1$ but $P_2$ has no length constraint,
and {\sc Edge-Disjoint $(= k_1, \ge k_2)$-Paths} requires that $|P_1| = k_1$ and $|P_2| \ge k_2$. \\

\noindent {\bf Related Work.} {\sc Edge-Disjoint $(L_1,L_2)$-Paths} has been studied 
under the framework of classical complexity. 
Ohtsuki~\cite{ohtsuki1980two}, Seymour~\cite{seymour1980disjoint}, 
Shiloah~\cite{shiloach1980polynomial}, and Thomasssen~\cite{thomassen19802} independently 
gave polynomial-time algorithms for {\sc Edge-Disjoint $(\le \infty, \le \infty)$-Paths}.
Tragoudas and Varol~\cite{tragoudas1997computing} proved the NP-completeness of
{\sc Edge-Disjoint $(\le k_1, \le k_2)$-Paths},
and Eilam-Tzoreff~\cite{eilam1998disjoint} showed the NP-completeness of
{\sc Edge-Disjoint $(\le k_1, \le \infty)$-Paths} even when $k_1$ equals the $(s_1,t_1)$-distance.
For {\sc Edge-Disjoint $(L_1,L_2)$-Paths} with $L_1 = k_1$ or $\ge k_1$ 
(same for $L_2 = k_2$ or $\ge k_2$), we can easily establish its NP-completeness by 
reductions from the classical {\sc Hamiltonian Path} problem.

As for the parameterized complexity, there are a few results in connection with our
{\sc Edge Disjoint $(L_1,L_2)$-Paths}.
Golovach and Thilikos~\cite{golovach2011paths} obtained an $2^{O(kl)}m \log n$-time
algorithm
for {\sc Edge Disjoint Paths} when every path has length at most $l$.
For a single pair $(s,t)$ of vertices, Fomin et al.~\cite{fomin2014efficient}
gave the currently fastest $O(2.851^l m \log^2 n)$-time algorithm for finding
an $(s,t)$-path of length exactly $l$, if it exists.
For the problem of finding an $(s,t)$-path of length at least $l$,
Bodlaender~\cite{bodlaender1993linear} derived an $O(2^{2l} (2l)!n + m)$-time algorithm,
Gabow and Nie~\cite{gabow2008finding} designed an $l^l 2^{O(l)}mn\log n$-time algorithm,
and a recent FPT algorithm of Fomin et al.~\cite{fomin2014efficient} for cycles
can be adapted to yield a $8^{l + o(l)} m\log ^2 n$-time algorithm. \\

\noindent {\bf Our Contributions.} 
In this paper, we investigate the parameterized complexity of {\sc Edge-Disjoint $(L_1,L_2)$-Paths}
for the nine different length constraints and have obtained FPT algorithms for seven of them 
(see \reftb{summary} for a summary).

In particular, we use random partition in an interesting way to obtain FPT algorithms for
{\sc Edge-Disjoint $(= k_1, \le \infty)$-Paths} and {\sc Edge-Disjoint $(= k_1, \ge k_2)$-Paths}.
This is achieved by bounding the number of some special edges, called ``nearby-edges'',
in the two paths $P_1$ and $P_2$ by a function of $k_1$ and $k_2$ alone.
We also consider polynomial kernels and prove that all nines cases admit no polynomial kernel 
unless $NP \subseteq coNP/poly$. \\

\tbl{summary}{Running times of FPT algorithms for {\sc Edge-Disjoint $(L_1,L_2)$-Paths} with 
	length constraints $L_i \in \{ \le k_i, \;  = k_i, \;  \ge k_i, \; \le \infty\}$ for $i = 1,2$.
	Note that $r_1 = k_1 + k_2$, $r_2 = k^2_1 + 5k_2$, and $r_3 = k^2_2 + 5k_1$. }{
	\setlength{\extrarowheight}{4pt}{
		\begin{tabular}{|p{2.0cm}|p{2.5cm}|p{2.0cm}|p{2.0cm}|p{2.5cm}|}
			\hline
			\multicolumn{1}{|c|}{ Constraints } & \multicolumn{1}{c|}{ $|P_2| \le k_2$ } & 
			\multicolumn{1}{c|}{ $|P_2| = k_2$ } & \multicolumn{1}{c|}{ $|P_2| \ge k_2$ } & 
			\multicolumn{1}{c|}{ $\le \infty$ } \\
			\hline
			\multicolumn{1}{|c|}{ $|P_1| \le k_1$ } & 
			\multicolumn{1}{c|}{ $O(2.01^{r_1} m \log n)$ } & & 
			\multicolumn{1}{c|}{ \multirow{2}{*}{ $O(2.01^{r_2} m\log^3 n)$ }} & 
			\multicolumn{1}{c|}{ $O(2.01^{k^2_1} m \log n)$ } \\
			\cline{1 - 2}
			\cline{5 - 5}
			\multicolumn{1}{|c|}{ $|P_1| = k_1$ } & 
			\multicolumn{2}{c|}{ $O(5.71^{r_1} m \log^3 n)$ } & & 
			\multicolumn{1}{c|}{ $O(2.01^{k^2_1} m \log^3 n)$ } \\
			\hline
			\multicolumn{1}{|c|}{ $|P_1| \ge k_1$ } & 
			\multicolumn{2}{c|}{ $O(2.01^{r_3} m\log^3 n)$ }  & 
			\multicolumn{2}{c|}{ Open } \\
			\hline
		\end{tabular}
	}
	\vspace*{4pt}
}

\noindent {\bf Notation and Definitions.} All graphs in the paper are simple undirected connected graphs. 
For a graph $G$, we use $V(G)$ and $E(G)$ to denote 
its vertex set and edge set respectively,
and $n$ and $m$, respectively, are numbers of vertices and edges of $G$. 
For two vertices $s$ and $t$, the distance between $s$ and $t$ is denoted by $d(s, t)$. 

An instance $(I, k)$ of a parameterized problem $\Pi$ 
consists of two parts: an input $I$ and a parameter $k$. 
We say that a parameterized problem $\Pi$ is fixed-parameter tractable (FPT) if 
there is an algorithm solving every instance $(I, k)$ in time $f(k)|I|^{O(1)}$ 
for some computable function $f$. 
A kernelization algorithm for a parameterized problem $\Pi$ maps an instance $(I , k)$ 
in time polynomial in $|I| + k$ into a smaller instance $(I', k')$ 
such that $(I , k)$ is a yes-instance iff $(I', k')$ is a yes-instance 
and $|I'| + k' \le g(k)$ for some computable function $g$. 
Problem $\Pi$ has a polynomial kernel if $g(k)$ is a polynomial function. 

For simplicity, we write $O(2.01^{f(k)})$ for $2^{f(k) + o(f(k))}$ 
as the latter is $O((2 + \epsilon)^{f(k)})$ for any constant $\epsilon > 0$ 
and we choose $\epsilon = 0.01$. 
In particular, $2^k k^{O(\log k)} = 2^{k + O(\log^2 k)} = O(2.01^k)$. 

In the rest of the paper, we present FPT algorithms for seven cases in Section 2, 
and show the nonexistence of polynomial kernels in Section 3. 
We conclude with some open problems in Section 4. 

\section{FPT algorithms}

Random partition provides a natural tool for finding edge-disjoint $(L_1,L_2)$-paths
in a graph $G$: We randomly partition edges of $G$ to form two graphs $G_1$ and $G_2$,
and then independently find paths $P_1$ in $G_1$ (resp., $P_2$ in $G_2$)
whose lengths satisfy $L_1$ (resp., $L_2$).

When our problem satisfies the following two conditions, the above approach
yields a randomized FPT algorithm and can typically be derandomized by universal sets.

\begin{enumerate}
	\item Whenever $G$ has a solution, the probability of ``$G_1$ contains required $P_1$ and
	$G_2$ contains required $P_2$'' is bounded above by a function of $k_1$ and $k_2$ alone.
	
	\item It takes FPT time to find required paths $P_1$ in $G_1$ and $P_2$ in $G_2$.
\end{enumerate}

Indeed, straightforward applications of the above method yield FPT algorithms for
{\sc Edge-Disjoint $(L_1,L_2)$-Paths} when $L_i \in \{ \le k_i, \; = k_i\}$
for $i = 1,2$.

\thm{}{
	{\sc Edge-Disjoint $(L_1,L_2)$-Paths} can be solved in $O(2.01^{k_1 + k_2} m\log n)$ time
	for $(L_1,L_2) = (\le k_1, \; \le k_2)$, and $O(5.71^{k_1 + k_2} m \log^3n)$ time for
	$(L_1,L_2) = (\le k_1, \; = k_2)$ or $(= k_1,\; = k_2)$.
}
\proof{
	Let $r = k_1 + k_2$. 
	We randomly color each edge by color 1 or 2 with probability 1/2 to
	define a random partition of edges.
	Denote by $G_i$, $i =1,2$, the graph consisting of edges of color $i$. 
	Then for all three cases of $(L_1,L_2)$, the probability that both $G_1$ and
	$G_2$ contain required paths is at least $1/2^r$ when
	{\sc Edge-Disjoint $(L_1,L_2)$-Paths} has a solution.
	
	We can use BFS starting from $s_i$ to determine 
	whether $G_i$ contains an $(s_i, t_i)$-path of length at most $k_i$ in time $O(m)$, 
	and an algorithm of Fomin et al.~\cite{fomin2014efficient} to determine 
	whether $G_i$ contains an $(s_i, t_i)$-path of length exactly $l$ in time $O(2.851^l m \log^2 n)$.
	Furthermore, we use a family of $(m, r)$-universal sets of size $2^r r^{O(\log r)} \log m$~\cite{naor1995splitters} for derandomization. 
	Therefore {\sc Edge-Disjoint $(L_1,L_2)$-Paths} can be solved in time 
	\[ 2^r r^{O(\log r)} \log m * m = 2^r r^{O(\log r)} m\log n = O(2.01^r m\log n)\]
	for $(L_1,L_2) = (\le k_1, \; \le k_2)$, and time 
	\[ 2^r r^{O(\log r)} \log m * (2.851^{k_1} + 2.851^{k_2}) m \log^2 n = O(5.71^r m\log^3 n)\]
	for $(L_1,L_2) = (\le k_1, \; = k_2)$ or $(= k_1,\; = k_2)$.
	\qed} \\

For other cases of $(L_1,L_2)$, a random edge partition of $G$ does not,
unfortunately, gurantee condition (1) because of the possible existence
of a long path in a solution.
To handle such cases, we will compute some special edges and then use
random partition on such edges to ensure condition (1).
For this purpose, we call a vertex $v$ a {\em nearby-vertex}
if $d(s_1, v) + d(v, t_1) \le k_1$, and call an edge a {\em nearby-edge}
if its two endpoints are both nearby-vertices.
We will show that there exists a solution where the number of nearby-edges
is bounded above by a polynomial in $k_1$ and $k_2$ alone, which enables us to
apply random partition to nearby-edges to ensure condition (1) and hence to
obtain FPT algorithms.
We note that such a clever way of applying random partition has been used by
Cygan et. al~\cite{cygan2014parameterized} in obtaining an Eulerian graph by deleting at most $k$ edges.

In the next two subsections, we rely on random partition of nearby-edges
to obtain FPT algorithms to solve {\sc Edge-Disjoint $(L_1,L_2)$-Paths}
for the following four cases of $(L_1,L_2)$:
$(\le k_1, \le \infty), (= k_1, \le \infty), (\le k_1, \ge k_2)$ and $(= k_1, \ge k_2)$.

\subsection{One short and one unconstrained}

In this subsection, we use random partition on nearby-edges to obtain FPT algorithms
for {\sc Edge-Disjoint $(L_1,L_2)$-Paths} when $(L_1,L_2)$ is $(\le k_1, \le \infty)$
or $(= k_1, \le \infty)$.
To lay the foundation of our FPT algorithms, we first present the following
crucial property on the number of nearby-edges in a special solution.
Recall that a nearby-vertex $v$ satisfies $d(s_1, v) + d(v, t_1) \le k_1$ and
both endpoints of a nearby-edge are nearby-vertices.

\lem{disjoint-paths-2}{
	Let $(s_1, t_1)$ and $(s_2, t_2)$ be two pairs of vertices in a graph $G = (V, E)$, 
	$P_1$ an $(s_1, t_1)$-path of length at most $k_1$, 
	and $P_2$ a minimum-length $(s_2, t_2)$-path edge-disjoint from $P_1$. Then 
	\begin{enumerate}
		\item all edges in $P_1$ are nearby-edges, and
		
		\item $P_2$ contains at most $(k_1 + 1)^2$ nearby-edges.
	\end{enumerate}
}
\proof{
	Statement 1 is obvious and we focus on Statement 2. 
	
	For a vertex $v$ in $P_2$, 
	we say that $v$ is a {\em $P_1$-near vertex} if there is a vertex $u$ in $P_1$ such that
	$G$ contains a $(u, v)$-path of length at most $k_1 / 2$ that is edge-disjoint from $P_1$.
	We call $v$ a {\em $u$-near vertex} when we want to emphasize the endpoint $u$, 
	and refer to such a $(u, v)$-path as a {\em $P_1$-near $(u, v)$-path}. 
	
	Let $v^*$ be a nearby-vertex in $P_2$. 
	Since $d(s_1, v^*) + d(v^*, t_1) \le k_1$, 
	there is an $(s_1, v^*)$-path or a $(t_1, v^*)$-path of length at most $k_1/2$. 
	As $s_1$ and $t_1$ are vertices of $P_1$, $v^*$ must be a $P_1$-near vertex. 
	Therefore each nearby-vertex in $P_2$ is a $P_1$-near vertex,
	and we bound the number of $P_1$-near vertices to prove this lemma. 
	
	Suppose to the contrary that $P_2$ contains at least $(k_1 + 1)^2 + 1$ $P_1$-near vertices. 
	Then by pigeonhole principle, there exists a vertex $u$ in $P_1$ 
	that has at least $k_1 + 2$ $u$-near vertices. 
	Sort these vertices along $P_2$ from $s_2$ to $t_2$. 
	Let $v_1$ and $v_2$ be the first and last vertex respectively. 
	Then the $(v_1, v_2)$-section of $P_2$ has length at least $k_1 + 1$. 
	Let $W$ be the $(v_1, v_2)$-walk concatenating the $P_1$-near $(u, v_1)$-path 
	and the $P_1$-near $(u, v_2)$-path. 
	Then $W$ contains at most $k_1$ edges and is edge-disjoint from $P_1$ by the definition of $P_1$-near path. 
	So we can replace the $(v_1, v_2)$-section by $W$ to obtain an $(s_2, t_2)$-walk 
	that contains an $(s_2, t_2)$-path shorter than $P_2$, contradicting to the minimality of $P_2$. 
	Therefore $P_2$ contains at most $(k_1 + 1)^2$ $P_1$-near vertices and thus nearby-vertices, 
	which implies that $P_2$ contains at most $(k_1 + 1)^2$ nearby-edges. 
	\qed} \\

The above lemma lays the ground for an FPT algorithm based on random partition.
Let $\{E_1,E_2\}$ be a random partition of nearby-edges,
and construct $G_1 = G[E_1]$ and $G_2 = G - E(G_1)$.
Note that whenever $G$ admits a solution, it has a solution $(P_1,P_2)$ such
that $P_2$ is a minimum-length $(s_2,t_2)$-path edge disjoint from $P_1$.
Lemma 1 implies that $P_1$ is inside $G_1$ with probability $\ge 1/2^{k_1}$,
and $P_2$ is inside $G_2$ with probability $\ge 1/2^{(k_1 + 1)^2}$.
This ensures that, with probability $\ge 1/2^{k_1}$,
$G_1$ contains an $(s_1,t_1)$-path of length at most $k_1$ and,
with probability at least $1/2^{(k_1 + 1)^2}$,  $G_2$ contains an $(s_2,t_2)$-path.
Therefore with probability $\ge 1/2^{k_1 + (k_1 + 1)^2}$,
we will be able to find a solution for $G$ by finding an $(s_1,t_1)$-path
of length at most $k_1$ in $G_1$ and an $(s_2,t_2)$-path in $G_2$.
This paves the way for the following randomized FPT algorithm for
{\sc Edge-Disjoint $(\le k_1, \; \le \infty)$-Paths}.
Note that the algorithm also works for {\sc Edge-Disjoint $(= k_1, \; \le \infty)$-Paths} once we change ``length $\le k_1$'' to ``length $k_1$'' in Step 3. \\

{\bf Algorithm 1: }

\begin{enumerate}
	\item Find all nearby-edges in $O(m)$ time by two rounds of BFS, one from
	$s_1$ and the other from $t_1$.
	
	\item Randomly color each nearby-edge by color 1 or 2 with probability 1/2,
	and color all remaining edges of $G$ by color 2.
	Let $G_i$ ($i = 1,2$) be the graph consisting of edges of color $i$.
	
	\item Find an $(s_1,t_1)$-path $P_1$ of length $\le k_1$ in $G_1$, and an
	$(s_2,t_2)$-path $P_2$ in $G_2$.
	Return $(P_1,P_2)$ as a solution if both $P_1$ and $P_2$ exist,
	and return ``No'' otherwise.
\end{enumerate}

Algorithm 1 solves {\sc Edge-Disjoint $(\le k_1, \; \le \infty)$-Paths}
with probability $\ge 1/2^{k_1 + (k_1 + 1)^2}$ and runs in $O(m)$ time,
as the two tasks in Step 3 for $G_1$ and $G_2$ also take $O(m)$ time.
Let $m'$ be the number of nearby-edges and $r = k_1 + (k_1 + 1)^2$.
We can use $(m',r)$-universal sets to derandomize our algorithm, and obtain
a deterministic FPT algorithm running in time
\[ 2^{r}r^{O(\log r)} \log n * m' = O(2.01^{k_1^2} m \log n).\]

For {\sc Edge-Disjoint $(= k_1, \; \le \infty)$-Paths}, Step 3 takes more
time as it takes $O(2.851^{k_1} m \log^2 n)$ time to find
an $(s_1, t_1)$-path $P_1$ of length $k_1$.
Therefore our deterministic FPT algorithm for the problem takes time
\[ 2^{r}r^{O(\log r)} \log m' * 2.851^{k_1} m \log^2 n = O(2.01^{k_1^2} m\log^3 n). \]

\thm{disjoint-paths-none-1}{
	{\sc Edge-Disjoint $(\le k_1, \le \infty)$-Paths} and {\sc Edge-Disjoint $(= k_1, \le \infty)$-Paths} can be solved in time $O(2.01^{k_1^2} m\log n)$ and $O(2.01^{k_1^2} m \log^3 n)$ respectively.
}

\subsection{One short and one long}

Now we consider {\sc Edge-Disjoint $(L_1,L_2)$-Paths} when $(L_1,L_2)$ is
$(\le k_1, \ge k_2)$ or $(= k_1, \ge k_2)$.
The main difficulty lies in the possibility that one path may be long, and we
overcome this obstacle by the following lemma similar to \refl{disjoint-paths-2}
to upper bound the number of nearby-edges in a special solution.
Again, the lemma enables us to use random partition on nearby-edges to
obtain FPT algorithms for both cases.

For an $(s_1, t_1)$-path $P$, a {\em $P$-valid $(s_2, t_2)$-path} is an $(s_2, t_2)$-path that is edge-disjoint from $P$ and has length at least $k_2$.

\lem{disjoint-paths-3}{
	Let $(s_1, t_1)$ and $(s_2, t_2)$ be two pairs of vertices in a graph $G = (V, E)$, 
	$P$ an $(s_1, t_1)$-path of length at most $k_1$, and 
	$Q$ a $P$-valid $(s_2, t_2)$-path of minimum length. Then
	
	\begin{enumerate}
		\item all edges in $P$ are nearby-edges, and
		
		\item at most $k^2_1 + 3k_1 + 2k_2$ edges of $Q$ are nearby-edges.  
	\end{enumerate}
}
\proof{
	Statement (1) is obvious and we focus on Statement (2). 
	
	For path $Q$, let $Q[s_2]$ denote the section containing the first $k_2 + 1$ vertices, 
	and $Q[t_2]$ the section containing the remaining vertices. 
	We show that $Q[t_2]$ contains at most $k^2_1 + 3k_1 + k_2$ nearby-edges,
	which implies Statement (2) as the remaining part of $Q$, i.e., $Q[s_2]$, has $k_2$ edges.
	
	Let $v$ be a vertex in $Q[t_2]$.
	We say that $v$ is a {\em $P$-near vertex} (resp., {\em $Q[s_2]$-near vertex}) 
	if there is a vertex $u$ in $P$ (resp., $Q[s_2]$) such that
	$G$ contains a $(u, v)$-path of length at most $k_1 / 2$ that is edge 
	disjoint from $P$ and vertex-disjoint from $Q[s_2]$ (except $u$).
	We refer to such a $(u, v)$-path as a {\em $P$-near $(u,v)$-path} 
	(resp., {\em $Q[s_2]$-near $(u,v)$-path}).
	
	Consider a nearby-vertex $v$ in $Q[t_2]$.
	Since $d(s_1, v) + d(v, t_1) \le k_1$, $G$ contains either an $(s_1,v)$-path or a 
	$(t_1,v)$-path of length at most $k_1 / 2$.
	If this path contains a vertex $v^*$ of $Q[s_2]$ 
	such that the $(v, v^*)$-section is edge-disjoint from $P$ and vertex-disjoint from $Q[s_2]$ except $v$, 
	then $v$ is a $Q[s_2]$-near vertex, and otherwise $v$ is a $P$-near vertex.
	Therefore all nearby-vertices in $Q[t_2]$ are $P$-near or $Q[s_2]$-near vertices,
	and we will put an upper bound on the number of nearby-vertices 
	in $Q[t_2]$ by limiting the numbers of $P$-near and $Q[s_2]$-near vertices.
	
	We can use the proof of \refl{disjoint-paths-2} to show that 
	{\em $Q[t_2]$ contains at most $(k_1 + 1)^2$ $P$-near vertices},
	and we now prove that
	{\em $Q[t_2]$ contains at most $k_1 + k_2 - 1$ $Q[s_2]$-near vertices}.
	
	Suppose to the contrary that $Q[t_2]$ contains at least $k_1 + k_2$ $Q[s_2]$-near vertices.
	Let $v_i$ denote the $i$-th $Q[s_2]$-near vertex in $Q[t_2]$ 
	when we travel along $Q[t_2]$ from its other endpoint to $t_2$.
	Let $h = \lceil{k_1 / 2}\rceil + 1$, and denote by $P_h$ a $Q[s_2]$-near $(u_h, v_h)$-path
	for some vertex $u_h$ in $Q[s_2]$.
	Denote by $v_q$ the first $Q[s_2]$-near vertex in $Q[t_2]$ when we travel along $P_h$ 
	from $u_h$ to $v_h$.
	
	Since all internal vertices of the $(u_h, v_q)$-section $S_1$ of $P_h$ is vertex-disjoint from $Q$,
	we can replace $Q[u_h,v_q]$ by $S_1$ to obtain an $(s_2,t_2)$-path $Q^*$
	(see Figure \ref{fig:cases} for illustration).
	Clearly, $Q^*$ is edge-disjoint from $P$ as both $Q$ and $S_1$ are edge-disjoint from $P$. 
	We show in two cases that $k_2 \le |Q^*| < |Q|$ to contradict the minimality of $Q$.
	
	Note that the $(v_q, v_h)$-section $S_2$ of $P_h$ is vertex-disjoint from $Q[s_2]$ and 
	edge-disjoint from $P$. 
	It follows that if the $(v_q, v_h)$-section $S'_2$ of $Q[t_2]$ is longer
	than $S_2$, we can replace $S'_2$ in $Q$ by $S_2$ to obtain an 
	$(s_2,t_2)$-walk $W$ that is edge-disjoint from $P$ and shorter than $Q$.
	Since the first $k_2$ vertices of $W$ are distinct vertices, we can obtain from $W$
	a $P$-valid $(s_2,t_2)$-path shorter than $Q$.
	Therefore we may assume that $|S_2| \ge |S'_2|$ by the minimality of $Q$.
	
	\begin{figure}[hbt]
		\centerline{
			\includegraphics[scale=1.00]{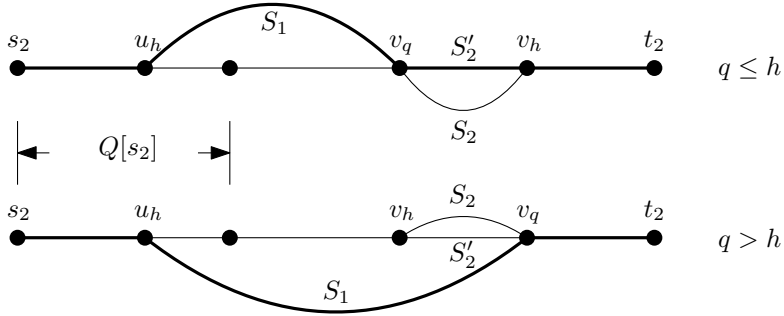}
		}
		\caption{Two cases for the intersections of $Q$ and $Q^*$. Note that $S_2$ may share internal vertices with $Q[t_2]$. }
		\label{fig:cases}
	\end{figure}
	
	{\bf Case 1:} $q \le h$.
	Clearly $|Q^*| \ge k_2$ as $Q[v_q,t_2]$ contains 
	more than $k_2$ $Q[s_2]$-near vertices.
	On the other hand, $|Q[u_h, v_h]| \ge h > |P_h|$ and $|S_2| \ge |S'_2|$.
	Therefore  
	\[ |S_1| = |P_h| - |S_2| < |Q[u_h, v_h]| - |S'_2| = |Q[u_h, v_q]| \]
	and hence $|Q^*| < |Q|$. 
	
	{\bf Case 2:} $q > h$.
	Clearly $|Q^*| < |Q|$ as $|S_1| \le k_1/2 < |Q[u_h,v_q]|$,
	and we show that $|Q^*| \ge k_2$.  
	Since $|S'_2| \le |S_2| \le k_1/2 - 1$, $S'_2$ contains at most $k_1/2$ $Q[s_2]$-near vertices.
	Therefore $q \le k_1$ and $Q[v_q, t_2]$ contains at least $k_2$ $Q[s_2]$-near vertices,
	implying $|Q^*| \ge k_2$.
	
	Therefore $Q[t_2]$ has at most $k_1 + k_2 - 1$ $Q[s_2]$-near vertices.
	Together with at most $(k_1 + 1)^2$ $P$-near vertices in $Q[t_2]$ and $k_2$ vertices in $Q[s_2]$,
	we conclude that $Q$ contains at most $k^2_1 + 3k_1 + 2k_2$ $P$-near and $Q[s_2]$-near vertices,
	and hence at most $k^2_1 + 3k_1 + 2k_2$ nearby-vertices/edges.
	\qed} \\


The above lemma enables us to obtain a randomized FPT for {\sc Edge-Disjoint $(\le k_1, \; \ge k_2)$} 
by replacing Step 3 of Algorithm 1 as follows:

\noindent {\bf Step 3:} Find an $(s_1,t_1)$-path $P_1$ of length $\le k_1$ in $G_1$, and an
$(s_2,t_2)$-path $P_2$ of length $\ge k_2$ in $G_2$.
Return $(P_1,P_2)$ as a solution if both $P_1$ and $P_2$ exist,
and return ``No'' otherwise.

By \refl{disjoint-paths-3}, the randomized algorithm solves {\sc Edge-Disjoint $(\le k_1, \; \ge k_2)$-Paths}
with probability $\ge 1 / 2^{k^2_1 + 4k_1 + 2k_2}$.  
Since an $(s_2,t_2)$-path $P_2$ of length $\ge k_2$ can be found in time $8^{k_2 + o(k_2)} m\log ^2 n$~\cite{fomin2014efficient} as mentioned earlier in the introduction, 
the two tasks in Step 3 takes $8^{k_2 + o(k_2)} m\log ^2 n$ time and thus 
the randomized algorithm runs in the same time. 
Let $m'$ be the number of nearby-edges and $r = k^2_1 + 4k_1 + 2k_2$.
We can use $(m',r)$-universal sets to derandomize our algorithm, and obtain
a deterministic FPT algorithm for {\sc Edge-Disjoint $(\le k_1, \; \ge k_2)$-Paths} running in time
\[ 2^{r}r^{O(\log r)} \log m' * 8^{k_2 + o(k_2)} m \log ^2 n = O(2.01^{k_1^2 + 5k_2} m \log^3 n).\]

For {\sc Edge-Disjoint $(= k_1, \; \ge k_2)$-Paths}, Step 3 needs to find
an $(s_1, t_1)$-path $P_1$ of length $k_1$ which takes $O(2.851^{k_1} m \log^2 n)$ time.
Therefore our deterministic FPT algorithm for the problem takes time
\[ 2^{r}r^{O(\log r)} \log m' * O(2.851^{k_1} m \log^2 n + 8^{k_2 + o(k_2)} m\log ^2 n) = O(2.01^{k_1^2 + 5k_2} m \log^3 n). \]

\thm{one-long-path}{
	Both {\sc Edge-Disjoint $(\le k_1, \ge k_2)$-Paths} and {\sc Edge-Disjoint $(= k_1, \ge k_2)$-Paths} 
	can be solved in time $O(2.01^{k_1^2 + 5k_2} m\log^3 n)$.
}

\section{Incompressibility}

Having obtained FPT algorithms, we are impelled to investigate the existence of polynomial
kernels for {\sc Edge-Disjoint $(L_1, L_2)$-Paths}.
Our findings are negative as we will show that, unless $NP \subseteq coNP/poly$,
the problem admits no polynomial kernel for all nine different cases of length
constraints $(L_1, L_2)$.

We start with relaxed-composition algorithms defined
by Cai and Cai~\cite{cai2014incompressibility},
which is a relaxation of composition algorithms introduced by
Bodlaender et al.~\cite{bodlaender2009problems} in their pioneer work on
the nonexistence of polynomial kernels.

\ddfn{relaxed-composition}{relaxed-composition~\cite{cai2014incompressibility}}{
	A relaxed-composition algorithm for a parameterized problem $\Pi$
	takes $w$ instances $(I_1, k), \dots , (I_w, k) \in \Pi$ as input and,
	in time polynomial in $\sum_{i = 1}^{w} |I_i| + k$, outputs an instance
	$(I, k) \in \Pi$ such that
	\begin{enumerate}
		\item $(I, k')$ is a yes-instance of $\Pi$ iff some $(I_i, k)$
		is a yes-instance of $\Pi$, and
		\item $k'$ is polynomial in $\max^w_{i = 1} |I_i| + \log w$.
	\end{enumerate}
}

Note that relaxed-composition algorithms relax the requirement in
composition algorithms~\cite{bodlaender2009problems} for parameter $k'$ from
polynomial in $k$ to polynomial in $\max^w_{i = 1} |I_i| + \log w$.
As observed by Cai and Cai~\cite{cai2014incompressibility},
the following important result is implicitly established in
Bodlaender et al.~\cite{bodlaender2009problems}.

\tthm{relaxed-composition}{\cite{bodlaender2009problems,fortnow2011infeasibility,bodlaender2014kernelization}}{
	If an NP-complete parameterized problem admits a relaxed-composition algorithm,
	then it has no polynomial kernel, unless $NP \subseteq coNP/poly$.
}


We also need the following polynomial parameter transformation (ppt-reduction in
short). 
\ddfn{ppt-reduction}{ppt-reduction~\cite{bodlaender2011kernel,cai2014incompressibility}}{
	A ppt-reduction from a parameterized problem $\Pi$ to another
	parameterized problem $\Pi'$ is an algorithm that, for input $(I, k) \in \Pi$, takes time polynomial
	in $|I| + k$ and outputs an instance $(I', k) \in \Pi'$ such that
	\begin{enumerate}
		\item $(I, k)$ is a yes-instance of $\Pi$ iff $(I', k')$ is a yes-instance of $\Pi'$, and 
		
		\item parameter $k'$ is bounded above by a polynomial of $k$.
	\end{enumerate}
}

\tthm{ppt-reduction}{\cite{bodlaender2011kernel}}{
	If there is a ppt-reduction from a parameterized problem $\Pi$ to another parameterized problem $\Pi'$, 
	then $\Pi'$ admits no polynomial kernel whenever $\Pi$ admits no polynomial kernel.
}

Now we show the nonexistence of polynomial kernels for seven easy cases. 
We first use relaxed-composition to show the nonexistence of polynomial kernels of 
{\sc $(s,t)$-$k$-Path} (resp., {\sc Long $(s, t)$-Path}) 
that are NP-complete problems of finding an $(s, t)$-path of length $k$ (resp., $\ge k$). 
Then we present ppt-reductions from these two problems to {\sc Edge-Disjoint $(L_1, L_2)$-Paths} problems. 

\lem{incompressibility-1}{
	Both {\sc $(s,t)$-$k$-Path} and {\sc Long $(s, t)$-Path} admit no polynomial kernel unless $NP \subseteq coNP/poly$.
}
\proof{
	Given $w$ instances of {\sc $(s,t)$-$k$-Path} with $s_i$ and $t_i$ 
	being the two terminal vertices of the $i$-th instance for $1 \le i \le w$, 
	we can relaxed-composite these $w$ instances into one instance 
	by identifying $s_i$ (resp.,  $t_i$) as one vertex for all $1 \le i \le w$. 
	Then, by \reft{relaxed-composition}, 
	{\sc $(s,t)$-$k$-Path} admits no polynomial kernel unless $NP \subseteq coNP/poly$. 
	By the same relaxed-composition, we can deduce that {\sc Long $(s, t)$-Path} admits no polynomial kernel unless $NP \subseteq coNP/poly$. 
} 

\thm{incompressibility-1}{
	{\sc Edge-Disjoint $(L_1, L_2)$-Paths} for
	$(L_1, L_2)$ being $(\le k_1, = k_2)$, $(\le k_1, \ge k_2)$, 
	$(= k_1, = k_2)$, $(= k_1, \le \infty )$, $(= k_1, \ge k_2)$, $(\ge k_1, \le \infty)$ or $(\ge k_1, \ge k_2)$,  admits no polynomial kernel unless $NP \subseteq coNP/poly$. 
}
\proof{
	Given an instance of {\sc $(s,t)$-$k$-Path}, 
	we construct an instance of {\sc Edge-Disjoint $(= k_1, \;  \le \infty)$-Paths} as following:
	\begin{enumerate}
		\item Set $s_1 = s$ and $t_1 = t$, and $k_1 = k$, 
		
		\item add new vertices $s_2$ and $t_2$, and edge $s_2t_2$.  
	\end{enumerate}
	
	The above reduction is clearly a ppt-reduction, 
	and thus {\sc Edge-Disjoint $(= k_1, \;  \le \infty)$-Paths} 
	admits no polynomial kernel unless $NP \subseteq coNP/poly$. 
	For the other six cases, similar ppt-reductions  
	from {\sc $(s,t)$-$k$-Path} or {\sc Long $(s,t)$-Path} will work. 
	\qed} \\

Now we consider the remaining two cases of length constraints $(\le k_1, \le k_2)$
and $(\le k_1, \le \infty)$.
Following our argument for the other cases, we can easily construct ppt-reductions from
the problem of determining whether $G$ contains an $(s,t)$-path of length at most $k$.
Unfortunately, this short path problem is solvable in polynomial time and
thus admits a polynomial kernel, which makes such ppt-reductions meaningless
for the purpose of proving the nonexistence of polynomial kernels.
In fact, these two cases are difficult to deal with, and we will design
delicate relaxed-composition algorithms to establish the nonexistence of
their polynomial kernels.

\thm{incompressibility-2}{
	Both {\sc Edge-Disjoint $(\le k_1, \le k_2)$-Paths} and {\sc Edge-Disjoint $(\le k_1, \le \infty)$-Paths} admit no polynomial kernel unless $NP \subseteq coNP/poly$. 
}
\proof{
	Let $(G^1, \le k_1, \le k_2), \dots, (G^w, \le k_1, \le k_2)$ be $w$ instances of {\sc Edge-Disjoint $(\le k_1, \le k_2)$-Paths}, and $n = \max^w_{i = 1} |V(G_i)|$. 
	Let $(s^i_1, t^i_1)$ and $(s^i_2, t^i_2)$ be the two pairs of vertices of the $i$-th instance for $1 \le i \le w$. 
	Assume that $w$ is a power of two, say $w = 2^d$. 
	Otherwise we can add some redundant no-instances to make $w$ a power of two. 
	
	We first show how to composite two instances into one instance, 
	which is the crucial step of our relaxed-composition. 
	Given the $i$-th instance and $j$-th instance, 
	we construct a new instance $(G', \le k'_1, \le k'_2)$ as following 
	(See \reff{incompressibility} for an illustration.): 
	\begin{enumerate}
		\item Create two pairs of vertices $(s'_1, t'_1)$ and $(s'_2, t'_2)$, 
		and four vertices $u_1, u_2, v_1$ and $v_2$. 
		
		\item Connect these new vertices with graph $G^i$ and $G^j$ as showed in \reff{incompressibility}, 
		where each dashed/dotted edge is a {\em short-path} of length one, and 
		each normal edge is a {\em long-path} of length $k_1 + 4$. 
		
		\item Denote by $G'$ the new graph and set $k'_1 = k_1 + 4$, $k'_2 = k_2 + 3(k_1 + 4) + 1$. 
	\end{enumerate}
	
	\begin{figure}[hbt]
		\centerline{
			\includegraphics[scale=1.00]{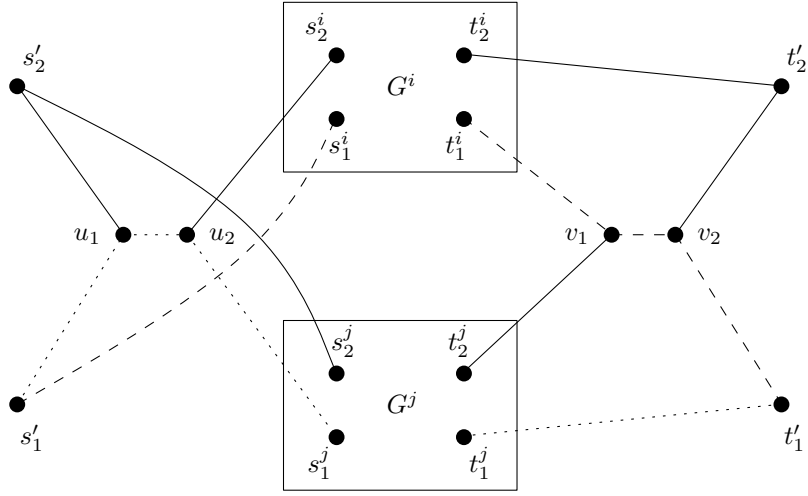}
		}
		\caption{The relaxed-composition for two instances. Here a dashed/dotted edge is a short-path of length one, and a normal edge is a long-path of length $k'_1 = k_1 + 4$.}
		\label{fig:incompressibility}
	\end{figure}
	
	We claim that $(G', \le k'_1, \le k'_2)$ is a yes-instance iff one of these two instances is a yes-instance.
	
	Suppose that one of these two instances has a solution. 
	Without loss of generality, assume that $(G^i, \le k_1, \le k_2)$ has a solution $(P_1, P_2)$. 
	Let $P'_1$ be the $(s'_1, t'_1)$-path concatenated by $P_1$ and the four dashed short-paths, 
	and $P'_2$ be the $(s'_2, t'_2)$-path going through $u_1, u_2, s^i_2$ and $t^i_2$, 
	whose $(s^i_2, t^i_2)$-section is $P_2$. 
	By the edge-disjointness between $P_1$ and $P_2$, $P'_1$ and $P'_2$ are edge-disjoint. 
	Furthermore, we have $|P'_1| \le k'_1$ and $|P'_2| \le k'_2$ as $|P'_1| \le k_1$ and $|P'_2| \le k_2$. 
	Then $(P'_1, P'_2)$ is a solution of $(G', \le k'_1, \le k'_2)$. 
	
	Conversely, suppose that $(P'_1, P'_2)$ is a solution of $(G', \le k'_1, \le k'_2)$. 
	Since $P'_1$ has length at most $k'_1 = k_1 + 4$, and each long-path has length $k_1 + 4$, 
	$P'_1$ contains either all dotted short-paths or dashed short-paths. 
	Assume that $P'_1$ contains all dotted short-paths. 
	(The argument is similar when $P'_1$ contains all dashed short-paths.)
	Then the $(s^j_1, t^j_1)$-section $P_1$ of $P'_1$ is an $(s^j_1, t^j_1)$-path in $G^j$ of length at most $k_1$. 
	Moreover, $P'_2$ must be an $(s'_2, t'_2)$-path going through 
	the $(s'_2, s^j_2)$-long-path $P_s$ and the $(t^j_2, v_1)$-long-path $P_t$.
	Since $d(v_1, t'_2) = k_1 + 5$, the $(s^j_2, t^j_2)$-section $P_2 \in G^j$ of $P'_2$ has length at most 
	\[|P'_2| - |P_s| - |P_t| - d(v_1, t_2) \le (k_2 + 3k_1 + 13) - 2(k_1 + 4) - (k_1 + 5) \le k_2.\]
	Then $(P_1, P_2)$ is a solution of $(G^j, \le k_1, \le k_2)$. 
	
	
	Now we are ready to present our relaxed-composition that contains $d = \log w$ iterations.
	In the $i$-th iteration, there are $2^{d - i + 1}$ instances and 
	we group these instances into $2^{d - i}$ pairs for $1 \le i \le d$. 
	For each pair, we composite them into one instance as presented above. 
	Finally, there remains only one instance which completes the relaxed-composition.
	Let $(\le k^i_1, \le k^i_2)$ be the length constraints after the $i$-th iteration for $0 \le i \le d$. 
	Note that $k^0_1 = k_1$ and $k^0_2 = k_2$. 
	The recursion relation for $k^i_1$ and $k^i_2$ is 
	\[ k^{i + 1}_1 = k^i_1 + 4 \mbox{ and } k^{i + 1}_2 = k^i_2 + 3k^{i + 1}_1 + 1, \] 
	as short-path and long-path respectively have length 1 and $k_1^{i + 1}$ in the $i$-th iteration. 
	We have $k^i_1 = k_1 + 4i$ and $k^i_2 = k_2 + (3k_1 + 1)i + 6i(i + 1)$ for $0 \le i \le d$. 
	
	Let $(G'', \le k''_1, \le k''_2)$ be the final instance, 
	where $k''_1 = k^d_1 = k_1 + 4d$ and $k''_2 = k^d_1 = k_2 + (3k_1 + 1)d + 6d(d + 1)$. 
	By above proof for the composition of two instances, 
	we can deduce that $(G'', \le k''_1, \le k''_2)$ has a solution 
	iff one of these $w$ instances has a solution. 
	Both $k''_1$ and $k''_2$ are polynomially bounded in $n + \log w$ as $d = \log w$. 
	This composition is a valid relaxed-composition. 
	Since {\sc Edge-Disjoint $(\le k_1, \le k_2)$-Paths} is NP-complete, by \reft{relaxed-composition}, 
	it admits no polynomial kernel unless $NP \subseteq coNP/poly$. 
	
	The relaxed-composition also holds if we discard the length constraint for the second path, i.e. 
	discard the length constraints $``\le k_2"$ and $``\le k'_2"$, 
	which yields that {\sc Edge-Disjoint $(\le k_1, \le \infty)$-Paths} 
	admits no polynomial kernel unless $NP \subseteq coNP/poly$.
	\qed} 

\section{Concluding Remarks}

We have obtained FPT algorithms to solve {\sc Edge-Disjoint $(L_1,L_2)$-Paths}
for seven of the nine different cases of length constraints $(L_1,L_2)$,
and also established the nonexistence of polynomial kernels for all nine cases,
assuming $NP \not\subseteq coNP/poly$.
However parameterized complexities of the remaining two cases are open. 

\pro{edge-disjoint}{
	Determine the parameterized complexities of {\sc Edge-Disjoint $(\ge k_1, \le \infty)$-Paths} and
	{\sc Edge-Disjoint $(\ge k_1, \ge k_2)$-Paths}. 
}

It is interesting to note that an FPT algorithm for
{\sc Edge-Disjoint $(\ge k_1, \ge k_2)$-Paths} will actually yield a new polynomial-time
algorithm to solve {\sc Edge-Disjoint Paths} for two pairs of terminal vertices (i.e.,
{\sc Edge-Disjoint $(\le \infty, \le \infty)$-Paths}).

We can consider vertex-disjoint paths, instead of edge-disjoint paths,
and form {\sc Vertex-Disjoint $(L_1,L_2)$-Paths} problems for nine different length
constraints $(L_1,L_2)$.
We note that it is straightforward to obtain FPT algorithms by color-coding
or random partition for the three cases of $(L_1,L_2)$
being $(\le k_1, \le k_2), (= k_1, \le k_2)$ or $(= k_1, = k_2)$,
but the remaining six cases seem much harder than their corresponding edge-disjoint
counterparts.

\pro{vertex-disjoint}{
	Determine the parameterized complexity of {\sc Vertex-Disjoint $(L_1,L_2)$-Paths}
	for the following six cases of $(L_1,L_2)$: \\
	$(\le k_1, \le \infty), (\le k_1, \ge k_2), (= k_1, \le \infty), (= k_1, \ge k_2),
	(\ge k_1, \le \infty)$ and $(\ge k_1, \ge k_2)$.
}

We note that structural properties similar to \refl{disjoint-paths-2} and
\refl{disjoint-paths-3}
seem not hold for vertex-disjoint paths with length constraints.
On the other hand, our proofs for the nonexistence of polynomial kernels for 
{\sc Edge-Disjoint $(L_1,L_2)$-Paths} also work for {\sc Vertex-Disjoint $(L_1,L_2)$-Paths}, and hence
{\sc Vertex-Disjoint $(L_1,L_2)$-Paths} admits no polynomial kernel
unless $NP \subseteq coNP /poly$ for all nine different cases of length constraints
$(L_1,L_2)$.

Finally, we can consider both edge-disjoint and vertex-disjoint paths with length constraints
for digraphs, which appear to be much harder than these problems on undirected graphs.

\pro{digraph}{
	For digraphs, determine the parameterized complexity of
	{\sc Edge-Disjoint $(L_1,L_2)$-Paths} and {\sc Vertex-Disjoint $(L_1,L_2)$-Paths}
	for various length constraints $(L_1,L_2)$.
}

\bibliographystyle{../splncs03}
\bibliography{disjoint-paths}

\end{document}